\documentclass[usenatbib]{mn2e}
\usepackage{graphicx}
\begin{document}

\pdfoutput=1 

\title[Observations and light curve solutions of six NSVS binaries]
{Ultrashort-period MS eclipsing systems. New observations and light curve solutions of six NSVS binaries}
\author[D. Dimitrov and D. Kjurkchieva]{Dinko P. Dimitrov$^{1}$\thanks{E-mail: dinko@astro.bas.bg; d.kyurkchieva@shu-bg.net} 
  and Diana P. Kjurkchieva$^{2}$\\
$^{1}$Institute of Astronomy and NAO, Bulgarian Academy of Sciences, Tsarigradsko shossee 72, 1784 Sofia\\
$^{2}$Department of Physics, Shumen University, 9700 Shumen, Bulgaria}

\date{Accepted 2015 January 19. Received 2015 January 12; in original form 2014 November 17}

\pagerange{\pageref{firstpage}--\pageref{lastpage}} \pubyear{2014}

\maketitle

\label{firstpage}

\begin{abstract}
We carried out photometric and low-resolution spectral observations of six eclipsing ultrashort-period binaries with MS components. The light curve 
solutions of the Rozhen observations show that all targets are overcontact systems. We found well-defined empirical relation ``period -- semi-major 
axis'' for the short-period binaries and used it for estimation of the global parameters of the targets. Our results revealed that NSVS~925605
is quite interesting target: (a) it is one of a few contact binaries with M components; (b) it exhibits high activity (emission in H$\alpha$ line, 
X-ray emission, large cool spots, non-Planck energy distribution); (c) its components differ in temperature by 700 K. All appearances of high magnetic 
activity and huge fillout factor (0.7) of NSVS~925605 might be assumed as a precursor of the predicted merging of close magnetic binaries. Another 
unusual binary is NSVS~2700153 which reveals considerable long-term variability.
\end{abstract}


\section{Introduction}

The short-period binaries with non-degenerate components are important objects for the astrophysics, especially for the understanding of the very late 
evolutional stages of the binaries connected with the processes of mass and angular momentum loss, merging or fusion of the stars, etc. But their 
structure and evolution remains unsolved problem in stellar astrophysics due to poor statistics. There are two reasons for this insufficiency.

Firstly, the period distribution of binaries reveals a very sharp decline in the number of short period systems below 0.27 days \citep{drake14}. 
Systems with periods around the short-period limit of 0.22 day \citep{rucinski92} were extremely rare: the fraction of ultrashort period objects 
was estimated to only 0.26 $\%$ of the total number of the large number of contact systems \citep{drake14}. Secondly, the faintness of the late 
short binaries makes them difficult targets for detailed study. As a result, although M-dwarfs form the most common stellar population
in our Galaxy \citep[$\sim$ 70 $\%$ by number, ][]{henry99}, their intrinsic faintness is barrier to study their binary characteristics.

Lately, a hypothesis appeared that the very short period low-mass binaries could merge via ``magnetic braking`` and if a substantial amount of mass 
remains in orbit around the primary, it would form a disk in which planets could be produced, particularly hot Jupiters \citep{martin11}.

Fortunately, the modern large stellar surveys during the last decade allowed to discover binaries with shorter and shorter periods 
\citep[][etc.]{rucinski07, pribulla09, weldrake04, maceroni97, dimitrov10, norton11, nefs12, davenport13, lohr14, qian14, drake14}.

This paper presents our observations and light curve solutions of six ultrashort-period binaries (whose periods are below 0.23 d).

\section{Selection of targets}

Using own approach \citep{dimitrov09} for searching for stellar variability we reviewed the NSVS database \citep{wozniak04} and found around 300 
ultrashort-period candidates with W UMa-type light curves. More than 100 of them were removed as probable $\delta$ Sct stars due to their small 
infrared colors (temperatures $T > 6000$ K). We carried out short test observations of the rest candidates and established that most of them were 
not variable stars. In such a way we managed to separate sample of about forty ultrashort-period candidates from the NSVS survey appropriate for 
follow-up observations at Rozhen observatory ($\delta > -10 \degr$). Our study of one of them (BX Tri) was already published \citep{dimitrov10}.

Table \ref{tab:targets} reveals the coordinates and magnitudes of six other targets from this sample: four of them are newly discovered eclipsing 
stars while two of them are identified with objects from the lists of ultrashort-period binaries of \citet{norton11} and \citet{lohr14} based on the 
SuperWASP photometric survey \citep{pollacco06}.

Besides photometric data for all six targets in the NSVS database we found also data of four of them in the SuperWASP archive {\citep{butters10}.
The periodogram analysis of these low-precise but numerous data allowed us to determine their ephemeris (Table \ref{tab:targets}). 
Figure \ref{fig:nsvs} presents the folded curves of the targets corresponding to their old photometric data.

\begin{table*}
\begin{minipage}[htp!]{\textwidth}
\caption{List of our ultrashort-period targets}
\label{tab:targets} \centering
\begin{small}
\renewcommand{\footnoterule}{}
\begin{tabular}{ccccccc}
\hline \hline
Target & NSVS ID  &   $\alpha$  &   $\delta$  & $R_{\rm {NSVS}}$ & Period & References \\
       &          &    [2000]   &     [2000]  & [mag]      & [d] & \\
\hline
 1 & 4484038 & 05 54 16.99 & +44 25 34.1 & 12.50 & 0.2185 & \citet{norton11} \\
 2 & 7179685 & 06 46 06.47 & +36 20 21.1 & 14.90 & 0.2097 & new \\
 3 & 4761821 & 08 01 50.03 & +47 14 33.8 & 13.10 & 0.2175 & \citet{norton11} \\
 4 & 2700153 & 13 43 51.43 & +63 04 22.8 & 12.48 & 0.2285 & new \\
 5 & ~925605 & 14 01 46.37 & +77 16 38.7 & 13.73 & 0.2176 & new \\
 6 & 8626028 & 21 16 26.81 & +25 17 36.2 & 13.78 & 0.2174 & new \\
\hline
\end{tabular}
\end{small}
\end{minipage}
\end{table*}

\section{Rozhen observations and data reduction}

The follow-up CCD photometry of the targets in $VRI$ bands was carried out with the three telescopes of Rozhen National Astronomical Observatory 
(Table \ref{tab:log1}). The 2-m RCC telescope is equipped with VersArray CCD camera (1340 $\times$ 1300 pixels, 20 $\mu$m/pixel, field of 5.35 
$\times$ 5.25 arcmin). The 60-cm Cassegrain telescope is equipped with FLI PL09000 CCD camera (3056 $\times$ 3056 pixels, 12 $\mu$m/pixel, field of 
17.1 $\times$ 17.1 arcmin). The 50/70-cm Schmidt telescope has FoV around 1$^\circ$ and is equipped with CCD camera FLI PL 16803, 
4096 $\times$ 4096 pixels, 9 $\mu$m/pixel size.

\begin{table}
\begin{minipage}[htp!]{\columnwidth}
\caption{Journal of the Rozhen photometric observations}
\label{tab:log1}
\centering
\begin{scriptsize}
\renewcommand{\footnoterule}{}
\begin{tabular}{ccrrr}
\hline \hline
Date  & Filter   & Exp. [s] &  Number & Telescope \\
\hline
\multicolumn{5}{c}{NSVS 4484038} \\
2010 Feb 24 & $VI$  & 120,60     & 16,16    & 60-cm \\
2010 Dec 08 & $VI$  & 120,60     & 90,90    & 60-cm \\
2011 Jan 08 & $VI$  & 120,90     & 87,86    & 60-cm \\
\multicolumn{5}{c}{NSVS 7179685} \\
2009 Sept 29 & $R$   & 120        & 40      & 60-cm \\
2009 Sept 30 & $R$   & 120        & 132     & 60-cm \\
2009 Oct 23 & $VRI$ & 120,120,120& 50,50,50 & 60-cm \\
2010 Mar 06 & $R$   & 120        & 95       & 60-cm \\
2010 Mar 12 & $I$   & 120        & 310      & 60-cm \\
2011 Jan 01 & $V$   & 40         & 434      &  2-m  \\
\multicolumn{5}{c}{NSVS 4761821} \\
2009 Nov 19 & $R$   & 120        & 50       & 60-cm \\
2011 Jan 01 & $VI$  & 120,60     & 103,103  & 60-cm \\
2011 Jan 06 & $VI$  & 120,120    & 85,85    & 60-cm \\
2011 Jan 13 & $VI$  & 120,120    & 18,16    & 60-cm \\
\multicolumn{5}{c}{NSVS 2700153} \\
2010 May  16 & $VI$  & 120,120    &  8,20    & 60-cm \\
2010 June 20 & $VI$  & 120,120    & 86,85    & 60-cm \\
2010 Nov 21  & $VI$  & 120,120    & 21,21    & 60-cm \\
2011 Jan 17  & $VI$  & 120,120    & 82,82    & Shmidt \\
2011 Jan 18  & $V$   & 120        & 125      & Shmidt \\
2011 Jan 19  & $I$   & 120        & 155      & Shmidt \\
\multicolumn{5}{c}{NSVS 925605} \\
2009 June 05 & $R$   & 120        & 100      & 60-cm \\
2009 June 06 & $R$   & 120        & 100      & 60-cm \\
2009 July 27 & $V$   & 120        & 91       & 60-cm \\
2009 Aug 26  & $VRI$ & 120,60,60  & 50,50,50 & 60-cm \\
2009 Aug 27  & $VRI$ & 120,60,60  & 60,60,60 & 60-cm \\
\multicolumn{5}{c}{NSVS 8626028} \\
2009 Aug 26 & $VRI$ & 120,60,60  & 50,50,50 & 60-cm \\
2009 Aug 27 & $VRI$ & 120,60,60  & 59,60,60 & 60-cm \\
\hline
\end{tabular}
\end{scriptsize}
\end{minipage}
\end{table}

Standard stars of \citet{landolt92} and standard fields of \citet{stetson00} were used for transition from instrumental system to standard photometric 
system.

The standard \textsc{IDL} procedures (adapted from \textsc{DAOPHOT}) were used for reduction of the photometric data. More than 5 standard stars were 
chosen in the  observed fields of each target by the requirement to be constant within 0.015 mag during the all observational runs and in all filters. 
Table \ref{tab:colors} presents their $V$, $B-V$, $V-R$, and $V-I$ determined by our observations (for the targets they correspond to 
phase 0.25) as well as their $J-K$ colors from the 2MASS catalog. 

Figures \ref{fig:1}-\ref{fig:6} present the follow-up Rozhen photometry of our targets. Table \ref{tab:phot} presents a sample of the Rozhen 
photometric data (the full table is available in the online version of the article, see Supporting Information).

We calculated times of the observed light minima by the method described in \citet{pribulla12}. Table \ref{tab:minima} contains their $HJD$(Min), 
Epoch, and $O-C$.

\begin{figure}
 \centering
 \includegraphics[width=\columnwidth,keepaspectratio=true]{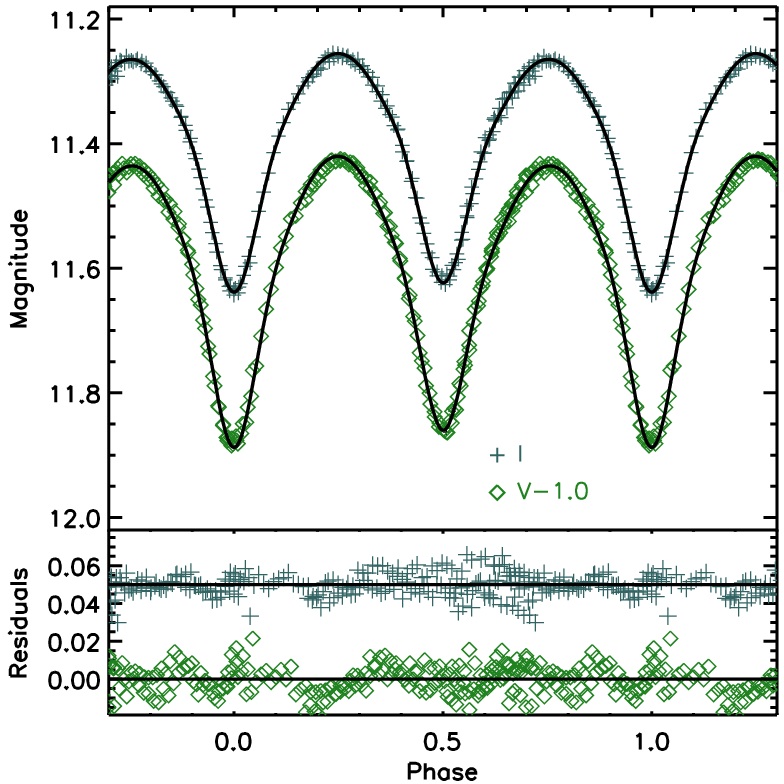}
 \caption{Top: the folded light curves of NSVS~4484038 and their fits (continuous lines); Bottom: the corresponding residuals. The photometric 
data in some filters are shifted vertically by different number for a better visibility. Color version of this figure is available in the online 
journal.}
 \label{fig:1}
\end{figure}

\begin{figure}
 \centering
 \includegraphics[width=\columnwidth,keepaspectratio=true]{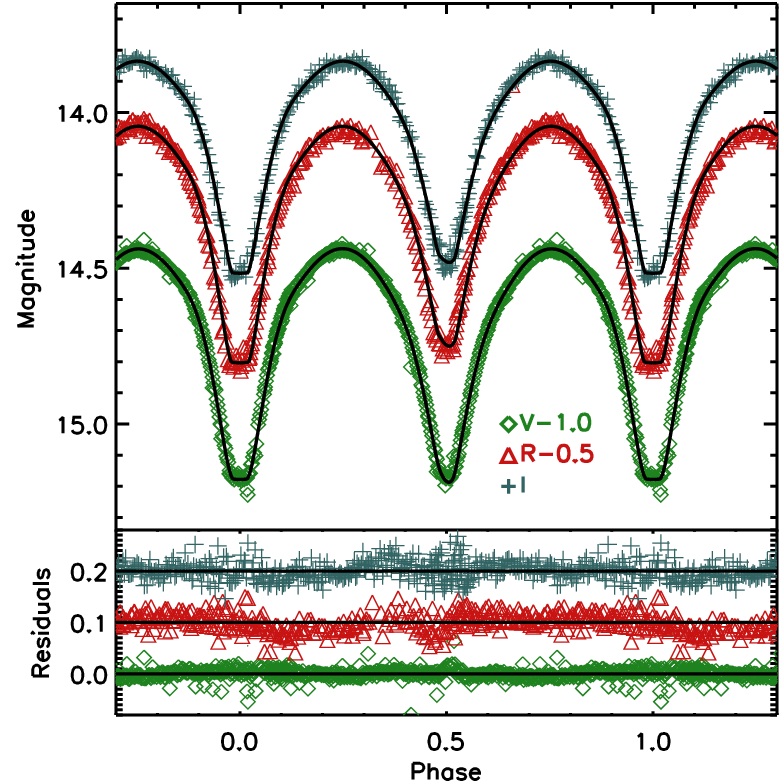}
 \caption{Same as Fig. \ref{fig:1} for NSVS~7179685.}
 \label{fig:2}
\end{figure}

\begin{figure}
 \centering
 \includegraphics[width=\columnwidth,keepaspectratio=true]{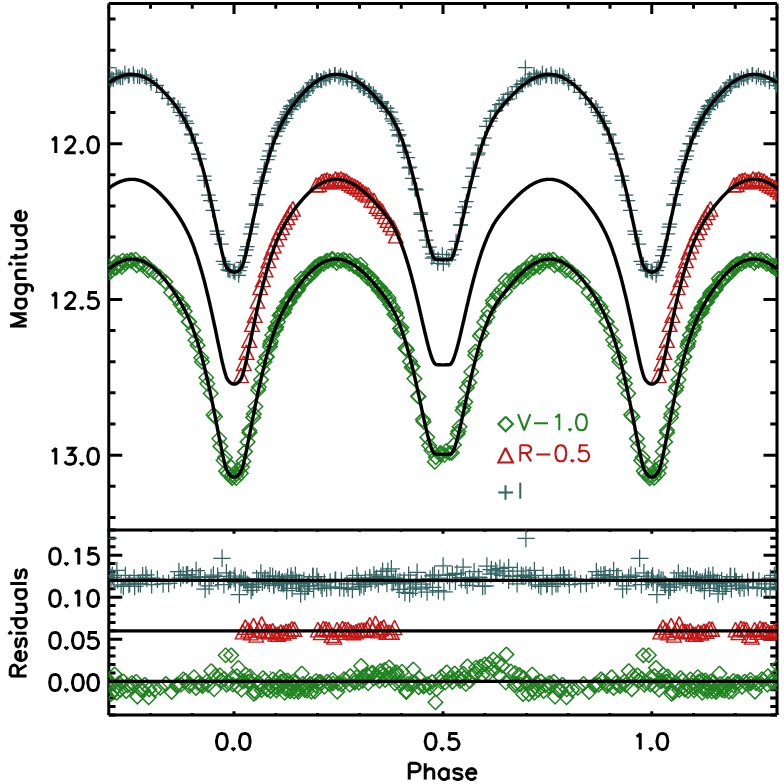}
 \caption{Same as Fig. \ref{fig:1} for NSVS~4761821.}
 \label{fig:3}
\end{figure}

\begin{figure}
 \centering
 \includegraphics[width=\columnwidth,keepaspectratio=true]{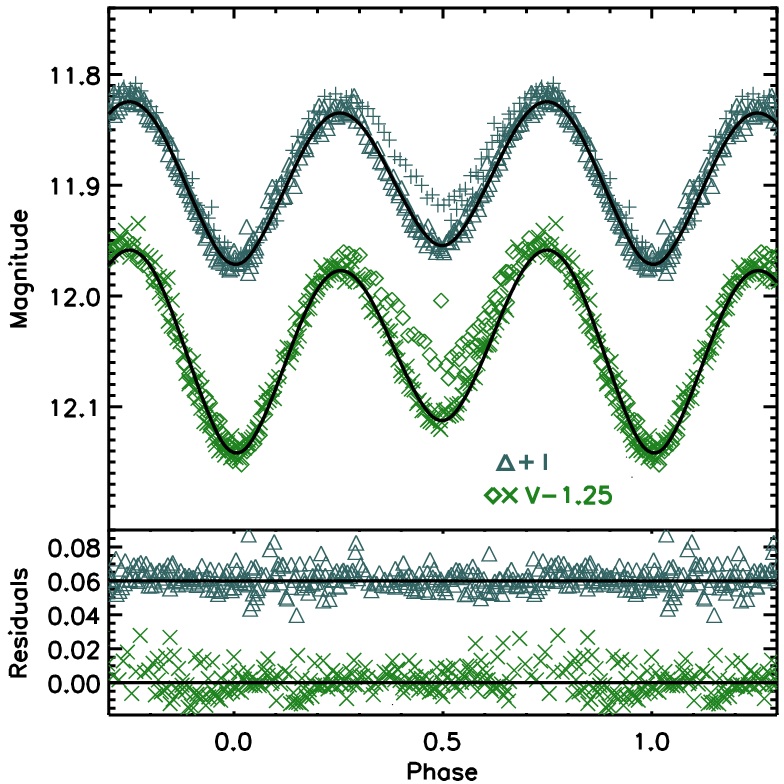}
 \caption{Same as Fig. \ref{fig:1} for NSVS~2700153.}
 \label{fig:4}
\end{figure}

\begin{figure}
 \centering
 \includegraphics[width=\columnwidth,keepaspectratio=true]{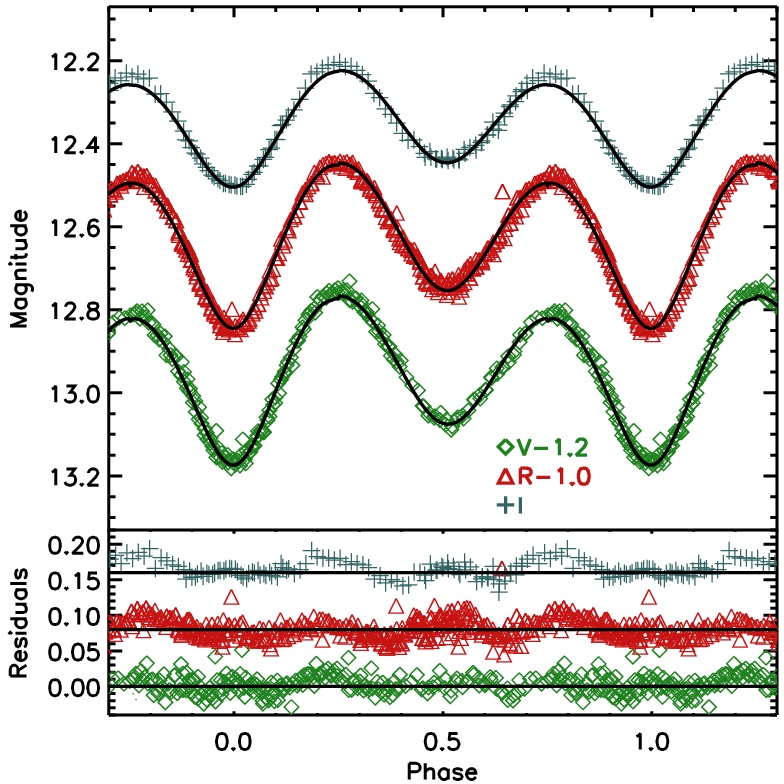}
 \caption{Same as Fig. \ref{fig:1} for NSVS~~925605.}
 \label{fig:5}
\end{figure}

\begin{figure}
 \centering
 \includegraphics[width=\columnwidth,keepaspectratio=true]{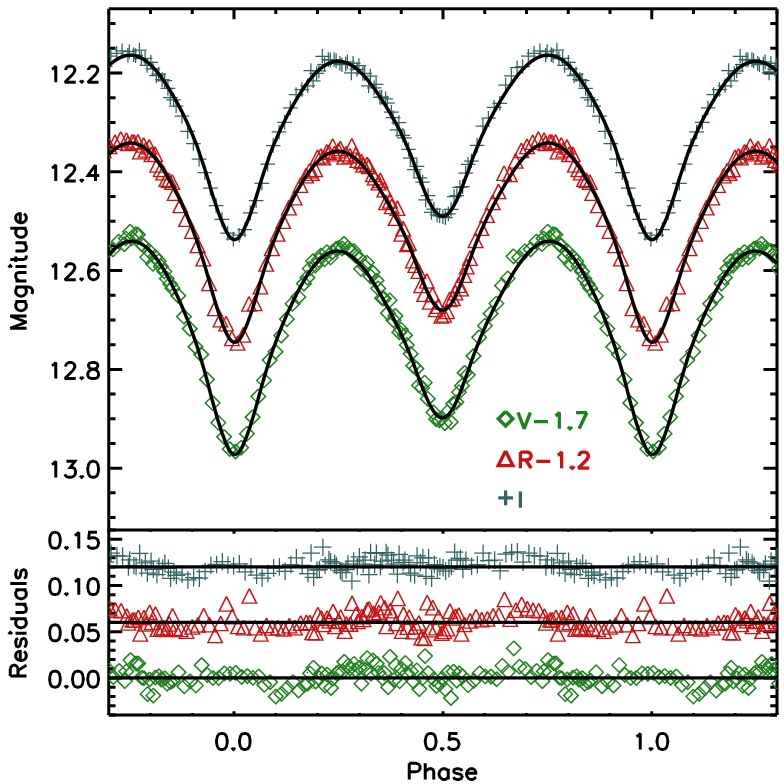}
 \caption{Same as Fig. \ref{fig:1} for NSVS~8626028.}
 \label{fig:6}
\end{figure}

We obtained low-resolution spectra (Fig. \ref{fig:spectra}) during 5 nights in Feb-May 2011 by the 2-m RCC telescope equipped with
focal reducer FoReRo-2. We used grism with 300 lines/mm that allows resolution 5.223 \AA/pixel in the range 5000-7000 \AA. These low-resolution 
spectra do not allow radial velocity measurement but are useful indicators of the temperature and stellar activity.

\begin{figure}
 \centering
 \includegraphics[width=0.95\columnwidth,keepaspectratio=true]{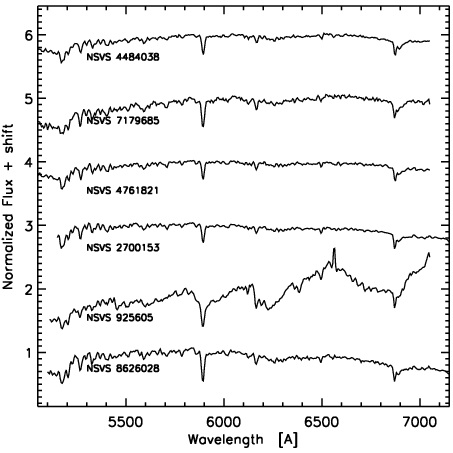}
 \caption{The low-resolution spectra of the targets around the H$\alpha$ line}
 \label{fig:spectra}
\end{figure}

\section{Light curve solutions}

The photometric data (Figs. \ref{fig:1}-\ref{fig:6}) implied that our targets are contact or overcontact systems that was expected for their
ultra-short orbital periods.

\begin{table*}
\begin{minipage}[htp!]{\textwidth}
\caption{The 2MASS color indices $J-K$ and corresponding mean temperatures $T_{\rm {m}}$ of the targets}
\label{tab:Tm}
\centering
\begin{scriptsize}
\renewcommand{\footnoterule}{}
\begin{tabular}{ccccccc}
\hline \hline
Target        & NSVS~4484038    & NSVS~7179685    &  NSVS~4761821   & NSVS~2700153    & NSVS~~925605    & NSVS~8626028    \\
\hline\
$J-K$         & $0.590\pm0.030$ & $0.798\pm0.033$ & $0.651\pm0.033$ & $0.635\pm0.037$ & $0.863\pm0.028$ & $0.769\pm0.028$ \\
$T_{\rm {m}}$ &    $4950\pm150$ & $   4040\pm190$ &    $4690\pm130$ &    $4730\pm200$ &    $3490\pm250$ &    $4200\pm180$ \\
\hline
\end{tabular}
\end{scriptsize}
\end{minipage}
\end{table*}

We carried out modeling of the Rozhen photometric data by the code \textsc{PHOEBE} \citep{prsa05} using the following procedure.

We determined in advance the mean temperatures $T_{\rm {m}}$ of the binaries (Table \ref{tab:Tm}) by their infrared color indices $(J-K)$ from the 
2MASS catalog and the calibration color-temperature of \citet{tokunaga00}. The "spectral" temperatures of the targets obtained by comparison of their 
low-resolution spectra (Fig. \ref{fig:spectra}) with those of standard stars with known temperatures, were almost the same as those of the values of 
$T_{\rm {m}}$ in Table \ref{tab:Tm}.

Firstly we assumed $T_{1}^{0}=T_{\rm {m}}$ and searched for solutions for fixed $T_{1}^{0}$ and $q^{0}=1$ varying the ephemeris, secondary temperature 
$T_{2}$, orbital inclination $i$ and potentials $\Omega_{1,2}$.

We adopted coefficients of gravity brightening 0.32 and reflection 0.5 appropriate for late stars. The linear limb-darkening coefficients were 
interpolated from the values of \citet{vanhamme93}. In order to reproduce the O'Connell effect we added cool spots on the components and varied their 
parameters: longitude $\beta$, latitude $\lambda$, angular size $\alpha$ and temperature factor $k=T_{\rm {sp}}/T_{1}$. To obtain a good fit in all 
colors a third light was necessary for some targets.

As a result of the first stage of the light curve solution we obtained initial values $T_{2}^{0}$, $i^{0}$ and $\Omega_{1,2}^{0}$ as well as the 
ephemeris and spot parameters for each target. Further we determined the mass ratio applying \textit{q-search} method. We calculated the normalized 
$\chi^{2}$ for two-dimensional grid ($i, \log q$) consisting of values of $i$ within the $10^\circ$-range around the value $i^{0}$ with step $0.5^{0}$ 
and $q$ values $0.15, 0.20,..., 0.95, 1.00, 1/0.95, 1/0.90,..., 1/0.20, 1/0.15$ ($T_{1}^{0}$, $T_{2}^{0}$ and spot parameters were fixed). In this way 
we obtained $\chi^{2}_{\rm {min}}(1)$ corresponding to the first approximation ($i^{1}, q^{1}$). Further this procedure was repeated around the pair 
($i^{1}, q^{1}$) for finer grid corresponding to 10 times smaller step of $q$ and 5 times smaller step of $i$. The obtained $\chi^{2}_{\rm {min}}(2)$ 
corresponded to the second approximation ($i^{2}, q^{2}$), etc. Thus we reach to the absolute minimum with $\chi^{2}_{\rm {min}}(abs) \sim 1$ 
corresponding to the final values ($i^{\rm {f}}, q^{\rm {f}}$).

The diagrams ($i, \log q$) exhibit that the solution ambiguity rapidly increases with the decreasing of the orbital inclination
(Fig. \ref{fig:q-search}). \citet{maceroni85} has reached to the same result by other procedure.

\begin{figure}
 \centering
 \includegraphics[width=0.95\columnwidth,keepaspectratio=true]{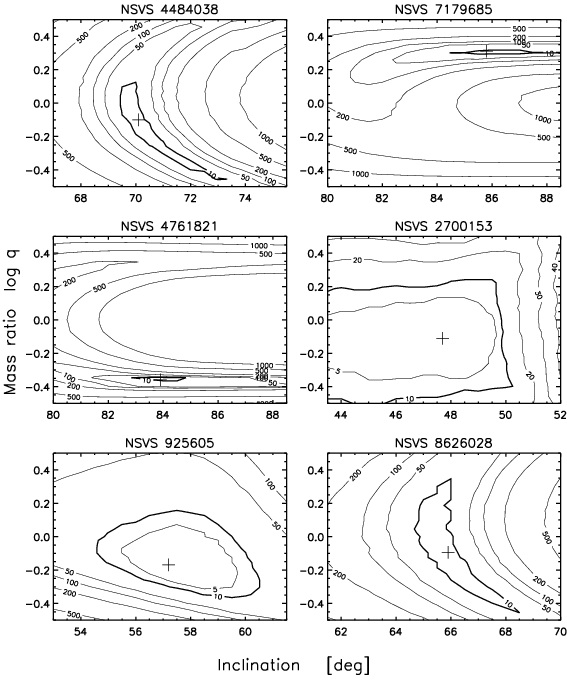}
 \caption{Diagrams $i$, $\log q$ with isolines corresponding to different values of the normalized $\chi^{2}$, the symbols $"+"$ mark the positions 
of $\chi^{2}_{\rm {min}}(abs)$}
 \label{fig:q-search}
\end{figure}

Further we searched for the best fit for fixed $i^{\rm {f}}$, $q^{\rm {f}}$, ephemerides and spot parameters, varying $T_{2}$, $\Omega_{1,2}$ 
and third light $l_{3}(V,R,I)$. Finally, we calculated $T_{1}$ and $T_{2}$. For this aim we developed the approach of \citet{coughlin11}

\begin{equation}
T_1=T_{\rm {m}} + \frac{\Delta T}{c+1}
\end{equation}
\begin{equation}
T_2=T_1 -\Delta T
\end{equation}
where $c=l_2/l_1$ and $\Delta T$ (difference of temperatures of the components) are determined from the \textsc{PHOEBE} solution.

Table \ref{tab:solution} contains the final values of all determined parameters: period $P$; initial epoch $HJD_{0}$; orbital inclination $i$; 
temperatures of the components $T_{1,2}$; potentials of the components $\Omega_{1,2}$; fill-out factor $FF$; spot parameters; third light contribution 
in the different colors $l_{3}(V,R,I)$. The errors of the parameters in Table \ref{tab:solution} are the formal \textsc{PHOEBE} values. 

The synthetic light curves corresponding to the parameters from Table \ref{tab:solution} are shown in Figs. \ref{fig:1}-\ref{fig:6} as continuous 
lines.

\begin{table*}
\begin{minipage}[htp!]{\textwidth}
\caption{Results from the light curve solutions}
\label{tab:solution} \centering
\begin{scriptsize}
\renewcommand{\footnoterule}{}
\begin{tabular}{clrrrrrr}
\hline \hline
\multicolumn{2}{c}{Parameter} & NSVS~4484038 & NSVS~7179685 & NSVS~4761821 & NSVS~2700153 & NSVS~~925605 & NSVS~8626028 \\
\hline
$P$               &  [d]  &    0.218493 &    0.209740 &    0.217513 &    0.228456 &    0.217629 &    0.217407  \\
                  &       &$\pm0.000001$&$\pm0.000001$&$\pm0.000001$&$\pm0.000001$&$\pm0.000001$&$\pm0.000001$ \\
$HJD_0$           &  [d]  & 5539.479711 & 5104.528662 & 5563.438743 & 5368.312230 & 4988.575181 & 5070.408406  \\
 +2450000         &       &$\pm0.000035$&$\pm0.000011$&$\pm0.000020$&$\pm0.000114$&$\pm0.000028$&$\pm0.000051$ \\
$i$               &[\degr]&     70.1 &     85.5 &     83.8 &     47.8 &     57.2 &     65.9  \\
                  &       & $\pm0.1$ & $\pm0.2$ & $\pm0.1$ &  $\pm0.4$&  $\pm0.2$&  $\pm0.1$ \\
$q=M_2/M_1$       &       &    0.792 &    2.128 &    0.432 &    0.775 &    0.678 &    0.805  \\
                  &       &$\pm0.002$&$\pm0.007$&$\pm0.001$&$\pm0.003$&$\pm0.003$& $\pm0.002$\\
$T_1$             &  [K]  &     5000 &     4100 &     4685 &     4785 &     3813 &      4318 \\
                  &       & $\pm43$  & $\pm32$  & $\pm32$  &$\pm51$   &  $\pm51$ &  $\pm27$  \\
$T_2$             &  [K]  &     4897 &     3979 &     4696 &     4689 &     3135 &      4095 \\
                  &       & $\pm41$  &  $\pm29$ &  $\pm29$ & $\pm40$  &  $\pm65$ &   $\pm22$ \\
$r_1$             &       &    0.407 &    0.326 &    0.464 &    0.409 &    0.474 &     0.418 \\
$r_2$             &       &    0.364 &    0.456 &    0.318 &    0.364 &    0.408 &     0.380 \\
$l_2/l_1$         &       &    0.727 &    1.731 &    0.479 &    0.709 &    0.221 &     0.592 \\	
$\Omega_1=\Omega_2$&      &    3.365 &    5.316 &    2.707 &    3.343 &    2.930 &     3.330 \\
                  &       &$\pm0.004$&$\pm0.006$&$\pm0.006$&$\pm0.007$&$\pm0.010$&$\pm0.004$ \\
$FF$              &       &   0.0857 &    0.193 &    0.136 &    0.071 &    0.702 &     0.207 \\
$Spot_1$          &       &  primary &  primary &    --    &    --    &  primary &   primary \\
$\lambda$         &[\degr]&      115 &      270 &    --    &    --    &      115 &       225 \\
$\beta$           &[\degr]&       90 &       90 &    --    &    --    &       63 &        70 \\
$\alpha$          &[\degr]&       13 &       15 &    --    &    --    &       16 &         9 \\
$\kappa$          &       &      0.9 &      0.9 &    --    &    --    &     0.84 &       0.9 \\
$Spot_2$          &       &     --   &      --  &    --    &    --    &secondary &   primary \\
$\lambda$         &[\degr]&     --   &      --  &    --    &    --    &      178 &       310 \\
$\beta$           &[\degr]&     --   &      --  &    --    &    --    &       90 &        90 \\
$\alpha$          &[\degr]&     --   &      --  &    --    &    --    &       18 &        11 \\
$T_{\rm {Spot2}}$ &  [K]  &     --   &      --  &    --    &    --    &      0.8 &       0.9 \\
$l_3(V)$          &       &    0.015 &    0.012 &    0.014 &    --    &    0.089 &     0.036 \\
                  &       &$\pm0.004$&$\pm0.003$&$\pm0.010$&    --    &$\pm0.012$&$\pm0.011$ \\
$l_3(R)$          &       &     --   &      --  &    --    &    --    &    0.078 &     0.048 \\
                  &       &     --   &      --  &    --    &    --    &$\pm0.010$&$\pm0.009$ \\
$l_3(I)$          &       &    0.112 &    0.007 &    --    &    0.069 &    0.293 &     0.056 \\
                  &       &$\pm0.003$&$\pm0.004$&    --    &$\pm0.008$&$\pm0.011$&$\pm0.005$ \\
\hline
\end{tabular}
\end{scriptsize}
\end{minipage}
\end{table*}

The parameter mass ratio deserves a special attention. Although its determination through analysis of the light-curve is an ambiguous approach 
compared with that by radial velocity solution, the rapid rotation of the components of the ultrashort-period binaries is serious obstacle to obtain 
precise spectral mass ratio from measurement of their highly broadened and blended spectral lines \citep{bilir05, dall05}. Contrariwise, tests have 
revealed that photometric mass ratios are more reliable than spectroscopic ones for totally eclipsing W~UMa-type stars and are sufficiently reliable 
for partially eclipsing systems \citep{maceroni96} because their eclipse depths depend strongly on the geometrical parameters and the mass ratio $q$. 
Hence, the obtained photometric $q$ values of our targets may be considered with a confidence.

\section{Analysis of the results}

The analysis of the light curve solutions led us to several conclusions.

\begin{description}
\item[(a)] All targets are overcontact (OC) binaries. The fill-out factor of NSVS~925605 is huge (Fig. \ref{fig:3D}).

\begin{figure}
 \centering
 \includegraphics[width=0.95\columnwidth,keepaspectratio=true]{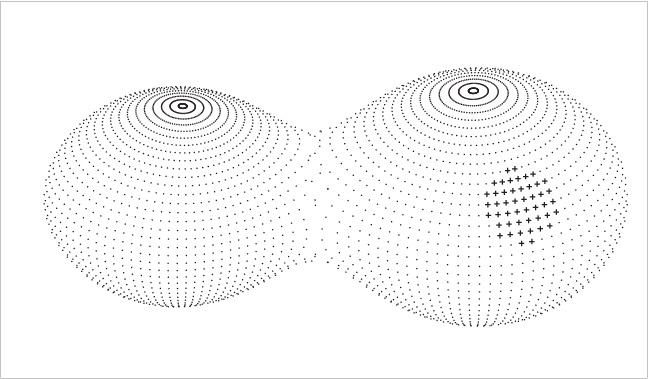}
 \caption{3D configuration of the strongly overcontact target NSVS~925605.}
 \label{fig:3D}
\end{figure}

\item[(b)]The mass ratios of four targets are within the narrow range 0.68--0.8 while NSVS~7179685 and NSVS~4761821 have components
which masses differ around 2 times. Just they are with total eclipses and thus with most precise photometric mass ratio.

\item[(c)] The differences between the temperatures of the components of the targets are up to 220 K, that is expected for overcontact systems. 
The only exception is NSVS~925605 which cool components differ by nearly 700 K (although the big fill-out factor).

\item[(d)] NSVS~7179685 is the only target which more massive star is the cooler component. We do not know overcontact binary with orbital period 
below 0.21 d with such a property.

\item[(e)] NSVS~~925605 is rare case of overcontact ultrashort period binary consisting of M dwarfs.

\item[(f)] NSVS~2700153 revealed considerable long-term variability during our observations (Fig. \ref{fig:4}): the secondary minimum was shallower 
than the primary one in June 2010, but they became almost equal in depth in Jan 2011. Parameters in Table \ref{tab:solution} correspond to the light 
curve solution from Jan 2011.

\item[(g)] NSVS~925605 shows high activity: emission in H$\alpha$ line (Fig. \ref{fig:spectra}) from its chromosphere; X-ray emission from the corona 
(it is the only X-ray source among our six targets identified as 1RXS J140139.8+771643); large cool photospheric spot. The considerable contribution 
of third light, especially in $I$ color (Table \ref{tab:solution}), cannot be explained by blurring because there is not nearby star within 1 arcsec 
from this target. The reason could be non-Planck distribution of its energy. All appearances of high magnetic activity and huge fillout factor of
NSVS~925605 might be interpreted in the light of the hypothesis of \citet{martin11} as a precursor of the merging of close magnetic binaries.
\end{description}

\section{Global parameters of the targets}

Due to lack of RV-curve solutions we had to use empirical relations for determination of the global characteristics of our targets. For this aim we 
built a diagram period--semi-major axis ($P,a$) on the base of 14 binaries with $P < 0.27$ d which have radial velocity solution and modern light 
curve solution (Table \ref{tab:stars}). Their distribution was approximated by parabola

\begin{equation}\label{equ:sma}
a = -1.154+14.633 \times P-10.319 \times P^2   \qquad [R\sun]
\end{equation}
\noindent where $a$ is in solar radii and $P$ in days. The standard deviation of the fit is 0.05 R$\sun$ and we assume this value as a mean 
error of $a$ for our targets with periods around 0.22 days.

Such a relation is expected for contact and overcontact binaries of MS late stars for which $R_1+R_2\approx a$ and $R \sim M^{k}$ where 
$k \approx 1$. Then the third Kepler law leads to $a \sim P$ that is similar to the almost linear dependence (\ref{equ:sma}) for periods of about 
0.22 d (Fig. \ref{fig:SMA}).

The ($P,a$) relation (\ref{equ:sma}) corresponds to the following relation ``period--mass'' for short-period binaries

\begin{equation}\label{equ:mass}
M = \frac {0.0134}{P^{2}} \left( -1.154+14.633 \times P-10.319 \times P^2 \right) ^{3} \, [M\sun]
\end{equation}

\noindent where $M$ is the total mass of the binary.

It is important to know is there a lower limit of the period-axis relation (\ref{equ:sma}). Very recently \citet{drake14} modeled binaries with 
extremely short periods by configurations consisting of white dwarf and dM star. They are stable non-accreting WD+MS systems which differ from the 
accretion induced variability in CV variables. We added these five systems (Table \ref{tab:stars}) on our diagram ($P,a$). It is interesting that 
their semiaxes, and respective masses, are almost the same (within the errors) while the orbital periods are quite different. The WD+dM models 
(blue asterisk) of four of these binaries fall away from the line approximating the relation ($P,a$) of binaries with MS components (red line). 
It is visible that the dependence of the semiaxis of these extremely short-period binaries on the period (blue continuous line) is considerably 
weaker. One of them, CSS J001242.4+130809, has 2 models: WD+dM and dM+dM (Table \ref{tab:stars}). Its dM+dM model as well as the longest-period 
target, CSS J090119.2+114254, fall just on the line period--axis of our MS binaries (Fig. \ref{fig:SMA}). This may imply that such configurations are 
more appropriate for these two targets. The low-resolution spectroscopy of the extremely short-period binaries \citep{drake14} revealed the presence 
of Balmer emission and Ca H+K emission typical for late dwarfs and strong local magnetic fields.

\begin{table*}
 \begin{minipage}[t]{\textwidth}
\caption{Semi-major axes of short-period binaries from spectral studies}
\label{tab:stars}
\renewcommand{\footnoterule}{}\centering
\begin{tabular}{l c l l l }
\hline \hline
Star         & $P_{\rm {orb}}$ & $M_{1}+M_{2}$& \multicolumn{1}{c}{$a$} &  References\\
\hline
CSS J090826.3+123648*      & 0.1392 & 0.80& 1.049& \citet{drake14} \\
CSS J111647.8+294602*      & 0.1462 & 0.75& 1.061& \citet{drake14} \\
CSS J081158.6+311959*      & 0.1562 & 0.70& 1.083& \citet{drake14} \\
CSS J001242.4+130809*      & 0.1641 & 0.80& 1.171& \citet{drake14} \\
CSS J001242.4+130809       & 0.1641 & 0.41& 0.937& \citet{drake14} \\
CSS J090119.2+114254*      & 0.1867 & 0.70& 1.222& \citet{drake14} \\
BX Tri                     & 0.1926 & 0.77$\pm$0.03 & 1.284$\pm$0.022 & \citet{dimitrov10} \\
BW3 V38                    & 0.1984 & 0.85$\pm$0.11 & 1.356$\pm$0.085 & \citet{maceroni04} \\
SDSS J001641-000925        & 0.1986 & 0.88$\pm$0.08 & 1.372$\pm$0.057 & \citet{davenport13} \\
GSC~1387-475               & 0.2178 & 0.94$\pm$0.03 & 1.492$\pm$0.019 & \citet{rucinski08} \\
CC Com                     & 0.2211 & 1.09$\pm$0.02 & 1.585$\pm$0.011 & \citet{pribulla07} \\
1SWASP J160156.04+202821.6 & 0.2265 & 1.43$\pm$0.06 & 1.761$\pm$0.033 & \citet{lohr14} \\
V523 Cas                   & 0.2337 & 1.13$\pm$0.04 & 1.663$\pm$0.025 & \citet{rucinski03} \\
RW Com                     & 0.2373 & 1.18$\pm$0.03 & 1.704$\pm$0.060 & \citet{pribulla09} \\
BI Vul                     & 0.2518 & 1.45& 1.899& \citet{maceroni96} \\
1SWASP J150822.80-054236.9 & 0.2601 & 1.62$\pm$0.10 & 2.013$\pm$0.065 & \citet{lohr14} \\
VZ Psc                     & 0.2612 & 1.46$\pm$0.06 & 1.950$\pm$0.040 & \citet{hrivnak95} \\
V803 Aql                   & 0.2634 & 1.58& 2.014& \citet{maceroni96} \\
FS Cra                     & 0.2636 & 1.51& 1.984& \citet{maceroni96} \\
44 Boo=i Boo               & 0.2678 & 1.47& 1.987& \citet{lu01} \\
\hline
\end{tabular}
\end{minipage}
Note: Stars with WD+dM model are marked with asterisk.
\end{table*}

We used the empirical relation (\ref{equ:sma}) to obtain the semi-major axes $a$ of our targets and further the masses $M_{1,2}$, radii $R_{1,2}$, and 
luminosities $L_{1,2}$ of their components in solar units; bolometric and visual absolute magnitudes $M_{\rm {bol}}$ and $M_V$ of the targets; 
distance $d$ in parsecs and orbital angular momenta $J_{\rm {orb}}$ \citep[by the expression of][]{popper77}. Table \ref{tab:param} contains the 
values of the obtained global parameters.

\begin{figure}
 \centering
 \includegraphics[width=0.95\columnwidth,keepaspectratio=true]{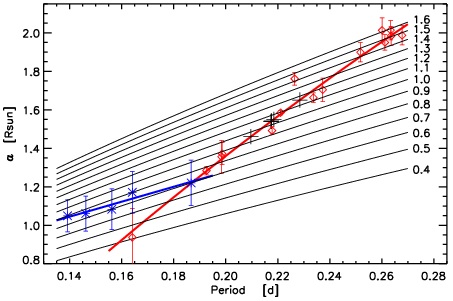}
 \caption{Diagram period -- semi-major axis for short-period binaries. The thin black lines represent the isolines of binary (total) mass.
The positions of the stars from Table \ref{tab:param} are marked by red diamonds while those of our targets are marked by black pluses.
The red line exhibits the empirical relation described by equation (\ref{equ:sma}). The blue line presents empirical relation
for WD+dM binaries, marked with blue asterisk. Color version of this figure is available in the online journal.}
 \label{fig:SMA}
\end{figure}

\begin{table*}
 \begin{minipage}[t]{\textwidth}
\caption{Global parameters of the targets}
\label{tab:param}
\renewcommand{\footnoterule}{}\centering
\begin{tabular}{cccccccccc}
\hline \hline
Name        & $a$       & $R_{1}$       & $M$       & $M_{1}$       & $L_{1}$         &$M_{\rm {bol}}$& $d$    & $J_{\rm {orb}}$ \\
            &           & $R_{2}$       &           & $M_{2}$       & $L_{2}$         & $M_{V}$       &        &            \\
\hline
NSVS 4484038& 1.55      & 0.63$\pm$0.02 & 1.01      & 0.56$\pm$0.06 & 0.23$\pm$0.02   & 5.73$\pm$0.11 & 200    & 0.151      \\
            & $\pm$0.05 & 0.57$\pm$0.02 & $\pm$0.10 & 0.45$\pm$0.04 & 0.17$\pm$0.02   & 5.928         & $\pm$12& $\pm$0.034 \\
NSVS 7179685& 1.461     & 0.48$\pm$0.02 & 0.95      & 0.30$\pm$0.03 & 0.06$\pm$0.01   & 6.68$\pm$0.13 & 428    & 0.118      \\
            & $\pm$0.05 & 0.67$\pm$0.02 & $\pm$0.09 & 0.65$\pm$0.06 & 0.10$\pm$0.01   & 7.284         & $\pm$26& $\pm$0.027 \\
NSVS 4761821& 1.541     & 0.71$\pm$0.02 & 1.04      & 0.72$\pm$0.07 & 0.23$\pm$0.02   & 6.00$\pm$0.10 & 248    & 0.137      \\
            & $\pm$0.05 & 0.49$\pm$0.02 & $\pm$0.10 & 0.32$\pm$0.03 & 0.11$\pm$0.01   & 6.397         & $\pm$12& $\pm$0.030 \\
NSVS 2700153& 1.650     & 0.67$\pm$0.02 & 1.15      & 0.65$\pm$0.06 & 0.22$\pm$0.02   & 5.81$\pm$0.11 & 267    & 0.190      \\
            & $\pm$0.05 & 0.60$\pm$0.02 & $\pm$0.11 & 0.50$\pm$0.05 & 0.16$\pm$0.02   & 6.060         & $\pm$14& $\pm$0.042 \\
NSVS  925605& 1.542     & 0.73$\pm$0.02 & 1.04      & 0.62$\pm$0.06 & 0.11$\pm$0.01   & 6.82$\pm$0.12 & 154    & 0.155      \\
            & $\pm$0.05 & 0.64$\pm$0.02 & $\pm$0.10 & 0.42$\pm$0.04 & 0.024$\pm$0.005 & 8.020         & $\pm$9 & $\pm$0.034 \\
NSVS 8626028& 1.540     & 0.63$\pm$0.02 & 1.04      & 0.57$\pm$0.06 & 0.13$\pm$0.01   & 6.40$\pm$0.10 & 184    & 0.159      \\
            & $\pm$0.05 & 0.57$\pm$0.02 & $\pm$0.10 & 0.47$\pm$0.04 & 0.09$\pm$0.01   & 7.904         & $\pm$9 & $\pm$0.035 \\
\hline
\end{tabular}
\end{minipage}
\end{table*}

The orbital angular momenta of the targets (Table \ref{tab:param}) are considerably smaller than those of the RS CVn binaries and detached 
systems which have $\log J_{\rm {orb}} \geq +0.08$. They are smaller also than those of the ordinary contact systems which have 
$\log J_{\rm {orb}} \geq -0.5$. The obtained $J_{\rm {orb}}$ values are bigger only than those of the shortest-period CVs of SU UMa type.

\section{Conclusions}

We found well-defined empirical relation ``period -- semi-major axis'' for the short-period binaries and used it for estimation of the global 
parameters of six ultrashort-period targets.

One of them, NSVS~925605, with huge fill-out factor and another peculiar characteristics (consisting of M components with high activity and non-Planck 
energy distribution) probably is at stage of merging.

Our results revealed that all six ultrashort-period targets with periods 0.20--0.23 days are overcontact systems. At the same time the short-period 
binaries BX Tri and BW3 V38 with periods below 0.20 have detached configurations. \citet{nefs12} spectroscopically confirmed another detached system
with a 0.18 day period containing an M-dwarf. It is worth to study which parameters determine the stellar configuration (detached, contact or 
overcontact) of the ultrashort-period binaries.

\section*{Acknowledgements}

The authors gratefully acknowledge observing grant support from the Institute of Astronomy and Rozhen National Astronomical Observatory, 
Bulgarian Academy of Sciences. The research was supported partly by funds of project RD-08-244 of Shumen University.

This research make use of the SIMBAD, Vizier, and Aladin databases, operated at CDS, Strasbourg, France, and NASA's Astrophysics Data System 
Abstract Service. This publication makes use of data products from the Two Micron All Sky Survey, which is a joint project of the University of 
Massachusetts and the Infrared Processing and Analysis Center/California Institute of Technology, funded by the National Aeronautics and Space 
Administration and the National Science Foundation. This research also has made use of the USNOFS Image and Catalogue Archive operated by the United 
States Naval Observatory, Flagstaff Station (http://www.nofs.navy.mil/data/fchpix/).

This research was based on data obtained with the telescopes at Rozhen National Astronomical Observatory, the Northern Sky Variability Survey 
(http://skydot.lanl.gov/nsvs/nsvs.php), and SuperWASP Public Archive first public data release DR1 (http://www.wasp.le.ac.uk/public/).

The authors are grateful to anonymous referee for the valuable notes and propositions.

\appendix
\section{Additional information}
\label{sec:appendix}

\begin{figure}
 \centering
 \includegraphics[width=0.99\columnwidth,keepaspectratio=true]{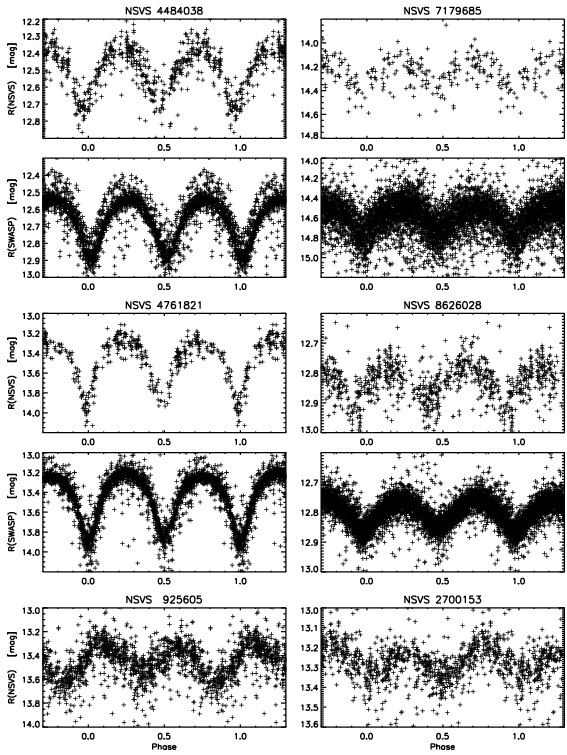}
 \caption{Folded light curves based on the NSVS and SuperWASP observations of the targets}
 \label{fig:nsvs}
\end{figure}

\begin{table}
\begin{minipage}[p]{\columnwidth}
\caption{Colors of the targets and their standard stars}
\label{tab:colors}
\centering
\begin{scriptsize}
\renewcommand{\footnoterule}{}
\begin{tabular}{l c c c c c c}
\hline \hline
 Star  &     ID      & $V$   & $B-V$ & $V-R$ & $V-I$ & $J-K$ \\
       &GSC1.2(2.3)  & [mag] & [mag] & [mag] & [mag] & [mag] \\
\hline
\multicolumn{7}{c}{NSVS 4484038} \\
Var & 2924-0179 & 12.43 & 0.71 & 0.25 & 1.17 & 0.59 \\
St1 & 2924-2161 & 12.86 & 1.00 & 0.40 & 1.31 & 0.69 \\
St2 & 2924-1407 & 12.68 & 1.23 & 0.70 & 1.93 & 1.08 \\
St3 & 2924-1435 & 13.23 & 0.60 & 0.16 & 0.78 & 0.37 \\
St4 & 2924-0641 & 12.02 & 1.87 & 0.68 & 1.88 & 1.07 \\
St5 & 2924-1202 & 11.79 & 0.86 & 0.13 & 0.69 & 0.34 \\
St6 & 2924-1060 & 12.94 & 0.97 & 0.46 & 1.40 & 0.75 \\
\multicolumn{7}{c}{NSVS 7179685} \\
Var & N8BM016247  & 15.44 &  0.45 &  0.91 &  1.61 &  0.80 \\
St1 & 2448-0250   & 13.42 &  0.65 &  0.71 &  1.20 &  0.63 \\
St2 & N8BN012958  & 13.63 &  0.35 &  0.48 &  0.77 &  0.34 \\
St3 & 2448-0995   & 13.29 &  1.34 &  0.77 &  1.42 &  0.74 \\
St4*& N8BM016279  & 15.19 &  0.46 &  0.72 &  1.23 &  0.64 \\
St5*& N8BM016265  & 15.64 &  0.28 &  0.71 &  1.29 &  0.69 \\
St6*& N8BM016250  & 15.38 & -0.20 &  0.43 &  0.69 &  0.24 \\
St7*& N8BM016262  & 15.20 &  0.08 &  0.55 &  0.96 &  0.48 \\
St8*& N8BM016256  & 15.82 &  0.10 &  0.55 &  0.95 &  0.44 \\
\multicolumn{7}{c}{NSVS 4761821} \\
Var & 3408-0735 & 13.37 & 1.06 & 0.76 & 1.60 & 0.65 \\
St1 & 3408-1475 & 11.58 & 0.82 & 0.74 & 1.44 & 0.49 \\
St2 & 3408-0617 & 13.51 & 0.66 & 0.61 & 1.15 & 0.33 \\
St3 & 3408-0431 & 12.59 & 0.69 & 0.83 & 1.72 & 0.74 \\
St4 & 3408-0459 & 13.26 & 0.54 & 0.63 & 1.14 & 0.30 \\
St5 & 3408-2017 & 13.54 & 0.73 & 0.66 & 1.18 & 0.31 \\
\multicolumn{7}{c}{NSVS 2700153} \\
Var & 4174-0776 & 13.19 & 0.82 & 0.71 & 1.37 & 0.64 \\
St1 & 4174-1064 & 12.42 & 0.71 & 0.39 & 1.00 & 0.43 \\
St2 & 4174-0970 & 12.99 & 0.68 & 0.42 & 0.95 & 0.43 \\
St3 & 4174-0772 & 13.29 & 0.42 & 0.41 & 0.83 & 0.30 \\
St4 & 4174-0696 & 12.07 & 0.53 & 0.43 & 0.83 & 0.31 \\
St5 & 4174-0764 & 13.76 & 0.35 & 0.46 & 0.96 & 0.41 \\
St6 & 4174-0526 & 14.03 & 0.46 & 0.37 & 1.02 & 0.46 \\
\multicolumn{7}{c}{NSVS  925605} \\
Var & 4558-0304 & 13.96 & 1.00 &  0.52 & 1.75 & 0.86 \\
St1 & 4558-0096 & 14.09 & 0.19 & -0.11 & 0.16 & 0.30 \\
St2 & 4558-0054 & 13.01 & 0.53 & -0.12 & 0.18 & 0.33 \\
St3 & 4558-0448 & 13.93 & 0.60 &  0.07 & 0.48 & 0.56 \\
St4 & 4558-0340 & 14.35 & 0.39 & -0.03 & 0.33 & 0.46 \\
St5 & 4558-0500 & 14.09 & 0.43 & -0.02 & 0.32 & 0.37 \\
St6 & 4558-2326 & 15.58 & 0.99 &  0.42 & 1.29 & 0.82 \\
\multicolumn{7}{c}{NSVS 8626028} \\
Var & 2190-2019 & 14.23 & 0.66 & 0.69 & 2.07 & 0.77 \\
St1 & 2190-1524 & 10.06 & 0.40 & 0.17 & 1.05 & 0.27 \\
St2 & 2190-1134 & 12.42 & 0.50 & 0.27 & 1.23 & 0.32 \\
St3 & 2190-1456 & 12.76 & 0.51 & 0.28 & 1.26 & 0.35 \\
St4 & 2190-0565 & 12.49 & 1.12 & 0.82 & 2.59 & 1.09 \\
St5 & 2190-2044 & 11.70 & 1.08 & 0.54 & 1.79 & 0.69 \\
\hline
\end{tabular}
\end{scriptsize}
\end{minipage}
* Standard stars used for the observations with the 2-m telescope
\end{table}

\begin{table}
\begin{minipage}[p]{\columnwidth}
\caption{Rozhen photometric data. This is a sample of the full table, which is available with the online version of the article 
(see Supporting Information). The number of the target is the same as in Table \ref{tab:targets}.}
\label{tab:phot}
\centering
\begin{scriptsize}
\renewcommand{\footnoterule}{}
\begin{tabular}{ccccc}
\hline \hline
Target & Filter & $HJD$  & Magnitude & Error  \\
   &   & [d] & [mag] & [mag] \\
\hline
1 & $V$ & 2455252.267213 & 12.8632 & 0.0059 \\
1 & $I$ & 2455252.269054 & 11.6127 & 0.0054 \\
1 & $V$ & 2455252.270188 & 12.8297 & 0.0059 \\
1 & $I$ & 2455252.272028 & 11.5911 & 0.0053 \\
... & ... &   ...   &   ...  & ... \\
2 & $R$ & 2455104.580844 & 14.5421 & 0.0175 \\
2 & $R$ & 2455104.582256 & 14.5606 & 0.0177 \\
2 & $R$ & 2455104.583680 & 14.5489 & 0.0177 \\
2 & $R$ & 2455104.585104 & 14.5417 & 0.0178 \\
... & ... &   ...   &  ...   & ... \\
\hline
\end{tabular}
\end{scriptsize}
\end{minipage}
\end{table}

\begin{table}
\begin{minipage}[htp!]{\columnwidth}
\caption{Times of the observed eclipse minima.} \label{tab:minima} \centering
\begin{scriptsize}
\renewcommand{\footnoterule}{}
\begin{tabular}{cclrr}
\hline \hline
 $HJD$(Min)  & Filter & Type of & Epoch & \multicolumn{1}{c}{$O-C$} \\
\multicolumn{1}{c}{[d]} & & minimum & & \multicolumn{1}{c}{[d]} \\
\hline
\multicolumn{5}{c}{NSVS~4484038} \\
2455539.479830 & V & Min I  &    0 &  0.000563 \\
2455539.479669 & I & Min I  &    0 & -0.000174 \\
2455539.589299 & V & Min II &    0 &  0.001572 \\
2455539.589183 & I & Min II &    0 &  0.001041 \\
2455570.287555 & V & Min I  &  141 & -0.001048 \\
2455570.287666 & I & Min I  &  141 & -0.000540 \\
2455570.397084 & V & Min II &  141 &  0.000236 \\
2455570.397060 & I & Min II &  141 &  0.000126 \\
\multicolumn{5}{c}{NSVS~7179685} \\
2455104.632827 & R & Min II &    0 & -0.003361 \\
2455105.472144 & R & Min II &    4 & -0.001659 \\
2455105.576676 & R & Min I  &    5 & -0.003271 \\
2455128.544335 & V & Min II &  114 &  0.002112 \\
2455128.543839 & R & Min II &  114 & -0.000253 \\
2455128.543751 & I & Min II &  114 & -0.000672 \\
2455128.648854 & V & Min I  &  115 &  0.000439 \\
2455128.648414 & R & Min I  &  115 & -0.001659 \\
2455128.648348 & I & Min I  &  115 & -0.001974 \\
2455262.463012 & R & Min I  &  753 &  0.000620 \\
2455268.335791 & I & Min I  &  781 &  0.000901 \\
2455268.440937 & I & Min II &  781 &  0.002217 \\
2455563.545264 & V & Min II & 2188 &  0.002918 \\
2455563.649773 & V & Min I  & 2189 &  0.001197 \\
\multicolumn{5}{c}{NSVS~4761821} \\
2455563.547579 & V & Min II &    0 &  0.000365 \\
2455563.546281 & I & Min II &    0 & -0.005603 \\
2455563.656336 & V & Min I  &    1 &  0.000367 \\
2455563.656667 & I & Min I  &    1 &  0.001889 \\
2455568.441653 & V & Min I  &   23 &  0.000509 \\
2455568.441644 & I & Min I  &   23 &  0.000468 \\
2455568.550426 & V & Min II &   23 &  0.000585 \\
2455568.550388 & I & Min II &   23 &  0.000411 \\
\multicolumn{5}{c}{NSVS~2700153} \\
2455368.313740 & V & Min I  &    0 &  0.006631 \\
2455368.313195 & I & Min I  &    0 &  0.004246 \\
2455368.425320 & V & Min II &    0 & -0.004959 \\
2455368.426162 & I & Min II &    0 & -0.001274 \\
2455579.518952 & V & Min II &  924 & -0.003699 \\
2455579.518799 & I & Min II &  924 & -0.004368 \\
2455580.548178 & V & Min I  &  929 &  0.001440 \\
2455580.660175 & V & Min II &  929 & -0.008325 \\
2455581.575046 & I & Min II &  933 & -0.003743 \\
2455581.690325 & I & Min I  &  934 &  0.000858 \\
\multicolumn{5}{c}{NSVS~~925605} \\
2454988.467268 & R & Min II &  -1 &  0.004147 \\
2454989.443816 & R & Min I  &   4 & -0.008639 \\
2454989.555544 & R & Min II &   4 &  0.004749 \\
2455040.368512 & V & Min I  & 238 & -0.010890 \\
2455070.297199 & V & Min II & 375 &  0.010704 \\
2455070.296609 & R & Min II & 375 &  0.007993 \\
2455070.296099 & I & Min II & 375 &  0.005650 \\
2455070.402749 & V & Min I  & 376 & -0.004296 \\
2455070.402535 & R & Min I  & 376 & -0.005280 \\
2455070.402487 & I & Min I  & 376 & -0.005500 \\
2455071.272856 & V & Min I  & 380 & -0.006176 \\
2455071.272715 & R & Min I  & 380 & -0.006824 \\
2455071.272511 & I & Min I  & 380 & -0.007761 \\
2455071.385397 & V & Min II & 380 &  0.010948 \\
2455071.384643 & R & Min II & 380 &  0.007483 \\
2455071.383811 & I & Min II & 380 &  0.003660 \\
\multicolumn{5}{c}{NSVS~8626028} \\
2455070.516938 & V & Min I  &   0 & -0.000789 \\
2455070.516287 & R & Min I  &   0 & -0.003783 \\
2455070.516218 & I & Min I  &   0 & -0.004101 \\
2455071.496252 & V & Min I  &   5 &  0.003730 \\
2455071.496096 & R & Min I  &   5 &  0.003013 \\
2455071.496085 & I & Min I  &   5 &  0.002962 \\
2455071.604069 & V & Min II &   5 & -0.000347 \\
2455071.603989 & R & Min II &   5 & -0.000715 \\
2455071.604153 & I & Min II &   5 &  0.000039 \\
\hline
\end{tabular}
\end{scriptsize}
\end{minipage}
\end{table}

\end{document}